\documentclass[seceq]{ptptex}

\usepackage{graphicx}

\title{Scaling Relation for Excitation Energy Under Hyperbolic Deformation}

\author{Hiroshi \textsc{Ueda},$^1$ Hiroki \textsc{Nakano},$^2$
Koichi \textsc{Kusakabe}$^1$ and Tomotoshi \textsc{Nishino}$^3$}

\inst{$^{1}$Department of Material Engineering Science, 
Graduate School of Engineering Science, Osaka University, Osaka 560-8531\\
$^{2}$Graduate School of Material Science, University of Hyogo,
Kamigori-cho, Hyogo 678-1297\\
$^{3}$Department of Physics, Graduate School of Science, Kobe University, Kobe 657-8501}

\recdate{June 15, 2010}%            %Editorial Office will fill in this.

\abst{
We introduce a one-parameter deformation for 
one-dimensional quantum lattice models, 
the hyperbolic deformation, where the scale of the local energy
is proportional to $\cosh \lambda j$ at the $j$th site. 
From the quantum-classical 
correspondence with a two-dimensional classical system, it is expected that
the deformation does not modify the ground state conspicuously, if  
there is a finite excitation gap. 
Under this situation, the excited quasi particle is weakly confined around 
the center of the system. We derive scaling relations among  the 
mean-square width  $\langle w^2_{~} \rangle$ of confinement, energy correction 
to the excitation gap $\Delta$, and deformation parameter $\lambda$. 
This scaling is useful for the extraction of the bulk excitation gap in the limit $\lambda = 0$. 
}

\begin{document}

\maketitle

\section{Introduction}

The ground state of an infinitely large quantum system that has a nonzero 
excitation gap is quite different from those of gapless systems 
in its correlation properties. Precise estimation of the excitation gap is therefore 
one of the theoretical interests in condensed matter physics. 
For one-dimensional (1D) quantum systems, the density matrix renormalization 
group (DMRG) method~\cite{White1,White2, Peschel, Schollwoeck} is often
used for the gap estimation, since it gives low-lying eigenvalues with high numerical 
precision. Since the target systems of DMRG are chiefly finite-size ones
under open boundary condition, finite-size scaling (FSS)~\cite{FSS1,FSS2} 
has been employed for the subtraction of the finite-size effect, which includes
boundary corrections.

Energies of the ground state and the first excited state are necessary for the
determination of the excitation gap.
The ground-state energy $E^{(0)}_L$ for a finite system of size $L$ ---  
typically up to the order of $L \sim 1000$ --- can be calculated numerically. 
Under the open boundary condition, 
an appropriate tuning of the system boundary 
is important for a rapid convergence of $E^{(0)}_L / L$ with 
respect to $L$, where $\varepsilon^{(0)}_{~} = \lim_{L \rightarrow \infty}^{~} 
E^{(0)}_L / L$ is the ground-state energy per site in the thermodynamic limit. 
To speak from a numerical viewpoint, there are several ways to obtain a faster
convergence of $\varepsilon^{(0)}_{~}$ with respect to $L$.  Since the number of
bonds of the $L$-site system is $L - 1$, $E^{(0)}_L / ( L - 1 )$ could be better 
than $E^{(0)}_L / L$. To observe the bond energy at the center of the system 
or to calculate $( E^{(0)}_{L+2} - E^{(0)}_L ) / 2$ is more efficient to avoid the 
boundary energy corrections. Further improvements can be achieved by a fine
tuning of interactions near the system boundary. For example, the 
introduction of weak bonds near the boundary~\cite{vekic:PRL71, vekic:PRB53} 
realizes a very fast decay of the boundary effect. To introduce sinusoidal modulation 
to the local energy scale is also useful for the suppression of the boundary 
effect.~\cite{Andrej:PTP122}

Precise estimation of the excitation gap $\Delta$ is more difficult 
than that of $\varepsilon^{(0)}_{~}$, even when the quasi-particle picture 
naturally holds for the elementary excitation. This is because the extrapolation process 
\begin{equation}
\Delta = 
\lim_{L \rightarrow \infty}^{~} \left( E^{(1)}_L - E^{(0)}_L \right) 
\end{equation}
requires energy of the first excited state $E^{(1)}_L$, 
which is easily affected by the reflection of 
the excited quasi particle at the system boundary. 
Such an effect may cause nonnegligible $L$-dependence 
on $E^{(1)}_L - E^{(0)}_L$. Thus, it is important to tune the 
interactions near the system boundary, so that the kinetic energy of the 
excited quasi particle rapidly (or regularly) 
converges to zero with respect to $L$.~\cite{Huse, Schollwock:PRB54, Nakano}
Generally speaking, it is not easy to perform the boundary tuning, 
which works efficiently for both ground and excited states. Obviously, the 
introduction of very weak bonds near the system boundary creates a {\it spurious} 
localized excitation near the boundary.

In this article, we consider a way of suppressing the boundary reflection effect, 
by weakly confining the excited quasi particle around the center of the system, 
preserving uniformity in the ground state as much as possible. 
To clarify the situation, let us consider the Hamiltonian 
\begin{equation}
{\hat H} = \sum^{~}_{j} {\hat h}^{~}_{j,j+1} + \sum^{~}_{j} {\hat g}_j
\end{equation}
on 1D lattice labeled by the site index $j$, 
where ${\hat h}^{~}_{j,j+1}$ and ${\hat g}^{~}_{j}$ denote neighboring 
interaction and on-site term, respectively.
We introduce a one-parameter deformation, the hyperbolic deformation, to the 
above Hamiltonian.~\cite{Ueda} The deformed Hamiltonian is written as
\begin{equation}
{\hat H}( \lambda ) = 
\sum^{~}_{j} \cosh \lambda j \,\, {\hat h}^{~}_{j,j+1} + 
\sum^{~}_{j} \cosh \lambda ( j - {\textstyle\frac{1}{2}} )  \,\, {\hat g}_j  \, ,
\end{equation}
where $\lambda$ is the deformation parameter. The ground state of this 
deformed system is expected to be nearly uniform, 
although the interaction strength is position-dependent. 
There is a geometrical reason for this uniformity, which we explain 
in the next section. 

Since the energy scale increases with respect to $| j |$ in 
Eq.~(1$\cdot$3), the excited quasi particle cannot 
reach the system boundary when $\lambda L$ is sufficiently large. This 
confinement gives a finite size correction to the energy of the first excited 
state $E_L^{(1)}( \lambda )$, where the correction $E_L^{(1)}( \lambda ) - 
E_L^{(1)}( 0 )$ comes from the uncertainty relation and its 
analytic form can be estimated. In order to detect this 
correction quantitatively for various gapped systems, 
we propose an effective one-particle 
Hamiltonian for the excited quasi particle. As we show in \S 3, there is a 
scaling relation among $L$, $E_L^{(1)}( \lambda ) - E_L^{(0)}( \lambda )$, 
and $\lambda$ under this quasi-particle picture. This scaling can be 
applicable for those systems that possess a nice quasi-particle picture, 
such as the Hubbard model~\cite{Kanamori,Gutzwiller,Hubbard} 
and integer-spin chains. Conclusions are summarized in the last section, 
and the applicability to the $S = 1$ antiferromagnetic Heisenberg spin chain 
is discussed.

\section{Uniformity under the hyperbolic deformation}

The real- or imaginary-time evolution of a 1D quantum system is related to a 
2D classical system through the so-called  quantum-classical 
correspondence.~\cite{Trotter,Suzuki,Baxter} Let us consider the 
correspondence for a case where a classical system is on 
a curved 2D space. We focus on the hyperbolic plane, which is a 2D space 
with a constant negative curvature. Suppose that there is a uniform 
action of the classical field on this curved surface. 
What does the corresponding 1D quantum Hamiltonian look like? 
Let us consider this problem from the viewpoint of the imaginary-time evolution. 

\begin{figure}
\begin{center} \includegraphics[width=50mm]{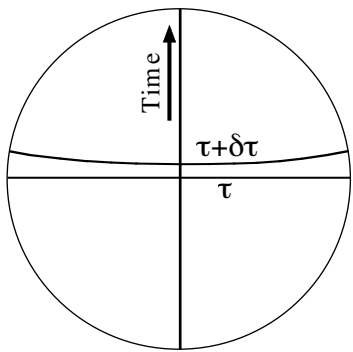} \end{center}
\caption{\label{hyperbolic}
Imaginary-time axis shown by the {\it vertical} line and {\it horizontal} 
equal-time curves 
on the hyperbolic plane drawn inside the Poincar\'e disc.
}
\end{figure}

Figure 1 shows the 2D hyperbolic plane drawn inside the Poincar\'e disc. 
All the geodesics are represented by arcs, which are perpendicular to the 
border circle, including straight lines that pass through 
the center of the disc. 
Let us choose the {\it vertical} line as the imaginary-time axis, 
where we denote the time by $\tau$. 
All the geodesics that are perpendicular to this imaginary-time 
axis can be regarded as equal-time curves. 
Consider a quantum state $| \Psi( \tau ) \rangle$ on an equal-time line. 
If the classical action in 
the lower half of the hyperbolic plane is uniform, and if there is no symmetry 
breaking such as dimerization, the state $| \Psi( \tau ) \rangle$ is also 
translationally invariant. 

Let us consider a tiny evolution from $\tau$ to $\tau + \delta \tau$ 
\begin{equation}
| \Psi( \tau + \delta \tau ) \rangle = {\hat  {\cal U}}
[ \delta \tau]  \,\, | \Psi( \tau ) \rangle \, ,
\end{equation}
where ${\hat {\cal U}}[ \delta \tau]$ represents the imaginary-time evolution. 
Although we have assumed that both $| \Psi( \tau ) \rangle$ and 
$| \Psi( \tau + \delta \tau ) \rangle$ are translationally invariant, 
${\hat {\cal U}}[ \delta \tau]$ is not. 
This fact can be understood geometrically by 
considering the distance between two points $( x, \tau )$ and 
$( x, \tau + \delta \tau )$ on the hyperbolic plane, 
where $x$ represents the spatial coordinate. 
The distance is an increasing function of $| x |$ as shown in Fig.~1, 
and can be written as $( \cosh \nu x ) \, \delta \tau$,
\footnote{The distance $( \cosh \nu x ) \, \delta \tau$ diverges in the limit 
$| x | \rightarrow \infty$. This divergence can be controlled by considering finite 
width $R$ and keeping $| x | < R$.}
where $\nu$ is a function of the scalar 
curvature of the hyperbolic plane. 
If it is possible to represent ${\hat {\cal U}}[ \delta \tau ]$ 
in the form $\exp\bigl( - \delta \tau { {\hat H}} \bigr)$,  
the corresponding Hamiltonian ${\hat H}$ is also position-dependent, and 
might be written with an integral 
\begin{equation}
{\hat H} = \int ( \cosh \nu x ) \, \hat{h}( x ) \, d x 
\end{equation}
of a local operator ${\hat h}( x )$. 
This is a rough sketch of the {\it hyperbolic deformation} 
of a quantum Hamiltonian in the continuous 1D space. 

Let us consider a construction of the hyperbolic deformation in the discrete space. 
We introduce lattice points at $x = a j$, where $a$ is the lattice constant, 
and $j$ the site index. (See Appendix A for the correspondence for the
one-particle case.) We also define a new dimensionless parameter 
$\lambda = \nu a$; we have
\begin{equation}
\cosh \nu x = \cosh \nu a j = \cosh \lambda j \, .
\end{equation}
A discrete analogue of ${\hat H}$ in Eq.~(2$\cdot$2) 
is then given by Eq.~(1$\cdot$3),  which is 
$
{\hat H}( \lambda )
=
\sum_j^{~} \cosh \lambda j   \,\, {\hat h}_{j, j+1}^{~}
+
\sum_j^{~} \cosh \lambda \bigl( j - {\textstyle \frac{1}{2}} \bigr)  \,\, {\hat g}_j^{~} \, .
$
When $\lambda = 0$, the Hamiltonian ${\hat H}( \lambda = 0 )$ 
coincides with the translationally invariant one in Eq.~(1$\cdot$1). 
From the construction of 
${\hat H}( \lambda )$,  it is expected that the ground state of this 
deformed system is nearly uniform, 
although the interaction strength in ${\hat H}( \lambda )$ 
is position-dependent. This assumption of the uniformity 
can be explicitly verified for  
classical statistical systems on hyperbolic lattices.~\cite{iharagi} 
Moreover, it was observed that the bond energy of the ground state of 
the $S = 1/2$ Heisenberg 
chain has very small position-dependence when the hyperbolic deformation 
is imposed.~\cite{Ueda}

\section{Finite-size scaling for the excitation energy} 

Compared with the ground state,  the effect of the hyperbolic deformation on  
the excited states is nontrivial. This is because quasi particles are subjected to 
the local interaction with the energy scale proportional to $\cosh \lambda j$. 
We analyze the behavior of the excited quasi particle using an effective  
one-particle model. Under the hyperbolic deformation, we can naturally 
expect that the effective Hamiltonian for the quasi particle also takes the form 
of the hyperbolic deformation 
\begin{equation}
{\hat H}_{\rm F}^{~}( \lambda ) = - t
\sum_j^{~} \cosh \lambda j  \left(
\, {\hat c}_j^\dagger {\hat c}_{j+1}^{~} +
\, {\hat c}_{j+1}^\dagger {\hat c}_j^{~} \right) 
- \mu \sum_j^{~} \cosh \lambda \bigl( j - {\textstyle \frac{1}{2}} \bigr)
\, {\hat c}_j^\dagger {\hat c}_j^{~} \, ,
\end{equation}
where $\hat{c}_j^\dagger$ and $\hat{c}_j^{~}$ 
represent, respectively, creation and annihilation of the quasi particle. 
They are either Fermionic or Bosonic, since
we consider the zero-particle vacuum $|0\rangle$ 
and one-particle states $|\Psi\rangle$ only, and therefore, 
the statistics of these operators does not matter in the following analysis. 
The letter `F' in ${\hat H}_{\rm F}^{~}( \lambda )$ denotes that 
${\hat H}_{\rm F}^{~}( \lambda )$ is quadratic, and is free from interaction terms.
For the moment, we assume that the system size is sufficiently large. 

We have denoted the hopping amplitude of the quasi particle by $t$. 
When the Hamiltonian is undeformed ($\lambda=0$), the parameter
$\mu$ can be interpreted as the chemical potential. 
Since we have introduced ${\hat H}_{\rm F}^{~}( \lambda )$ as an effective 
Hamiltonian for the gapped system, we consider the 
case when $\mu < -2t \cosh (\lambda/2)$. Under this condition,  
the ground state is trivially the zero-particle vacuum $| 0 \rangle$ with zero energy 
$E_{L}^{(0)}(\lambda )=0$, where $L$ is the system size as was introduced in \S 1,  
and the lowest excited state is the one-particle 
state $| \Psi \rangle$ with positive energy. 
It should be noted that we have assumed the translational invariance
of the ground state of ${\hat H}( \lambda)$ in Eq.~(1$\cdot$3), when we 
extract the effective Hamiltonian ${\hat H}_{\rm F}^{~}( \lambda )$.

Even for the toy model in Eq.~(3$\cdot$1), the distribution of the quasi 
particle for the lowest-energy one-particle state $| \Psi \rangle$ is nottrivial.  
The second term in Eq.~(3$\cdot$1) is relevant to the confinement effect. 
Introducing the wave function $\Psi_j = \langle j|\Psi\rangle 
=\langle 0|\hat{c}_j^{~}|\Psi\rangle$, we have 
\begin{equation}
E^{(1)}_{L}( \lambda ) \, \Psi_j^{~} =
- t \cosh \lambda j \, \Psi_{j+1}^{~}
- t \cosh \lambda \bigl( j - 1 \bigr) \,
\Psi_{j-1}^{~} 
- \mu \cosh \lambda \bigl( j - {\textstyle \frac{1}
{2}} \bigr) \, \Psi_j^{~} \, ,
\label{sch_eq}
\end{equation}
which is nothing but the one-particle Schr\"{o}dinger equation. (See Appendix A 
concerning the continuous limit.) Note that
the single-particle excitation energy when $L = \infty$ 
% $\Delta (\lambda)$ 
can be written as 
\begin{equation}
\Delta_{\infty}^{~}( \lambda) = 
E^{(1)}_{\infty}( \lambda ) - E^{(0)}_{\infty}( \lambda ) = 
E^{(1)}_{\infty}( \lambda ) \, ,
\end{equation}
which has the small-$\lambda$ limit
\begin{equation}
\Delta = \lim_{\lambda \rightarrow 0}^{~} \Delta_{\infty}^{~}( \lambda ) = - 2 t - \mu \, .
\label{Delta_3_4}
\end{equation}
Our aim is to find an efficient finite-size scaling, 
which captures this limit correctly. It is easy to solve 
Eq.~(\ref{sch_eq}) numerically when  the system size  $L$ is finite, up to $L$ of the order
of $10^2_{~} \sim 10^4_{~}$. Thus, what we have to do is to take the double limit 
\begin{equation}
\Delta = 
\lim_{\scriptstyle \lambda \rightarrow 0 \atop 
\scriptstyle L \rightarrow \infty}^{~} \,
\left[ E^{(1)}_L( \lambda ) - E^{(0)}_L( \lambda ) \right] = 
\lim_{\scriptstyle \lambda \rightarrow 0 \atop 
\scriptstyle L \rightarrow \infty}^{~} \, 
\Delta^{~}_L( \lambda ) 
\end{equation}
somehow by means of a scaling plot. 
% We have put subscript $L$ for each energy 
% variable in the above equation, such as $\Delta_L (\lambda)$. 
If there is an efficient finite-size scaling for $\Delta_L (\lambda)$, 
it could be useful for gap estimation of many-body models by means of the 
hyperbolic deformation in Eq.~(1$\cdot$3). As shown in the following, the double limit 
$\lambda\rightarrow 0$ and  $L\rightarrow \infty$ can be stably taken 
by keeping $(L+1)^2_{~} \lambda$ constant. 

\begin{figure}
\begin{center}
\includegraphics[width=75mm]{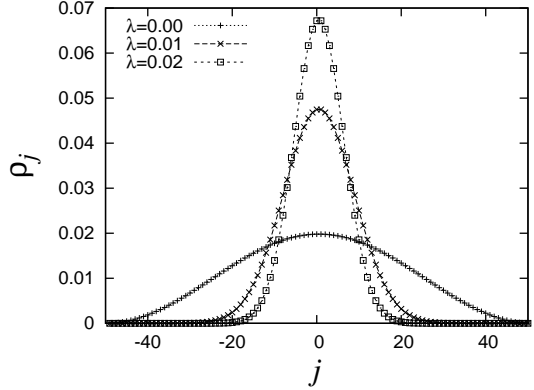}
\end{center}
\caption{
\label{density}
Site occupancy $\rho_j = \Psi^*_j \Psi_j$
under the conditions $L=100$, $t=1$, and $\mu = -3$. The gap $\Delta$ in Eq.~(3$\cdot$4) is 
equal to unity.
}
\end{figure}

Hereafter, we choose $t$ as the unit of energy, i.e., 
$t=1$. From the relation in Eq.~(\ref{Delta_3_4}), 
the parameter $\mu$ can be represented as 
$\mu = -2t - \Delta = - 2 - \Delta$. In the following, we specify
$\Delta$ instead of $\mu$ to simplify the formulations 
in the finite-size scaling. Figure 2 shows the occupancy 
$\rho_j=\Psi_j^*\Psi_j$ of the lowest-energy 
one-particle state when $\Delta = 1$ and $L=100$. 
We have labeled the sites with $j$ ranging from 
$j=-L/2+1$ to $j=L/2$. When $\lambda=0.01$ or $0.02$, 
the Gaussian-like distribution is observed around the center of the 
system $j=1/2$. 
The distribution can be characterized using the mean-square width 
\begin{equation}
\left \langle w^2_{~} \right \rangle = 
\sum^{L/2}_{j=-L/2+1}  \left( j - \frac{1}{2} \right)^{2}_{~} \, \rho_j^{~} \, .
\end{equation}

Consider the case where the parameter $\lambda$ is relatively small. As
we have examined in Fig.~2, confinement of the quasi particle appears when
$L$ is sufficiently large and $\lambda L \gg 1$ is satisfied. Under this condition, 
we can estimate the $\lambda$ dependence of 
$\langle w^2_{~} \rangle$ from the uncertainty relation. 
The typical energy to put a localized wave packet around 
the $j$th site is of the order of 
\begin{equation}
\Delta \, \cosh \lambda \left( j - {\textstyle \frac{1}{2}} \right) \sim  
\Delta \left[
1 + \frac{\lambda^2_{~}}{2} \left( j - {\textstyle \frac{1}{2}} \right)^2_{~}
\right] \, ,
\end{equation}
where we have truncated the Taylor series by the second order in $j-\frac{1}{2}$. 
Replacing $\left( j - \frac{1}{2} \right)^2_{~}$ by 
$\langle w^2_{~} \rangle$, 
we can roughly estimate the potential-like energy 
\begin{equation}
U \sim  \Delta + \Delta \, \frac{\lambda^2_{~} }{2} 
\left \langle w^2_{~} \right \rangle
\end{equation}
for the weakly confined state. From the uncertainty relation between 
position and momentum, 
% $\Delta x \Delta p \sim \hbar$, 
the kinetic-like energy 
$T = p^2_{~} / 2m$ can be estimated as 
\begin{equation}
T \sim \frac{b}{2 \left \langle w^2_{~} \right \rangle} \, ,
\end{equation}
where $b$ is a constant of the order of $t$. 
Minimizing the sum $E=U+T$, we obtain the relation 
\begin{equation}
\left \langle w^2_{~} \right \rangle \, \sim \, 
\frac{\sqrt{b}}{ \sqrt{\Delta} \, \lambda} 
\end{equation}
after a short calculation. Similarly, we can express the energy correction as
$E \, \sim \, \sqrt{b \Delta} \, \lambda \, $, which is caused by the confinement.

On the other hand, when $\lambda = 0$, the confinement effect disappears.
In this case, the quasi-particle wave function is exactly given by
\begin{equation}
\Psi^{~}_{j} = \sqrt{\frac{2}{L+1}} \, \cos \frac{j - \frac{1}{2}}{L+1} \, \pi  \, ,
\end{equation}
which represents a free  particle inside the area of width $L+1$. 
Therefore, $\langle w^2_{~} \rangle$ is 
of the order of $(L+1)^2_{~}$, where the proportional constant 
$c$ is calculated as 
\begin{equation}
c = 
\lim^{~}_{L \rightarrow \infty} \frac{\left \langle w^2_{~} \right \rangle}{( L+1 )^2_{~}} = 
\int^{1/2}_{-1/2} x^2_{~} \cos^2_{~} \pi x \,\, dx = 
\frac{1-6/\pi^2_{~}}{12} \, .
\end{equation}
The energy $\Delta_{L}^{~}(0) = E_{L}^{(1)}(0)$ is obtained as
\begin{equation}
\Delta_{L}^{~}(0) 
= -2t \cos \left( \frac{\pi}{L+1} \right) - \mu 
= -2 \cos \left( \frac{\pi}{L+1} \right) + 2 + \Delta  \, ,
\end{equation}
and therefore, the finite size correction to the 
true gap $\Delta \equiv \lim_{L \rightarrow \infty}^{~} 
\Delta_L^{~}( 0 )$ is derived as
\begin{equation}
\Delta_L^{~}( 0 ) - \Delta = 
\frac{\pi^2_{~} }{(L+1)^2_{~}} + 
\mathcal{O}\left( \frac{1}{(L+1)^4_{~}} \right) 
\, = \, \frac{\pi^2_{~} c }{\left \langle w^2_{~} \right \rangle} \, .
\end{equation}
\begin{figure}
\begin{center} \includegraphics[width=70mm]{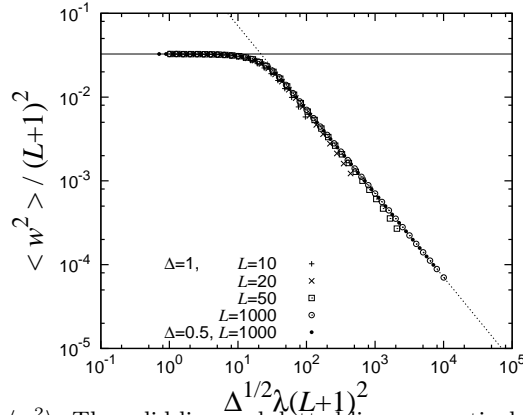} \end{center}
\caption{
Scaling plot for $\langle w^2_{~} \rangle$. 
The solid line and dotted line, respectively, correspond to 
$\langle w^2_{~} \rangle/(L+1)^2_{~} = c$ and 
$\langle w^2_{~} \rangle/(L+1)^2_{~} = \sqrt{b}/y$, 
where $c=(1-6/\pi^2_{~})/12$, $b=1/2$, and $y=\sqrt{\Delta}\lambda(L+1)^2_{~}$. 
}
\end{figure}

These observations suggest that $\left( \Delta_L^{~}( \lambda ) - \Delta \right) 
\left \langle w^2_{~} \right \rangle$ 
is a slowly varying function of $\lambda$, and that the scaling relation
\begin{equation}
\frac{\left \langle w^2_{~} \right \rangle}{( L + 1 )^2_{~}} = 
f\left[ \sqrt{\Delta} \, \lambda ( L + 1 )^2_{~} \right] 
\end{equation}
is satisfied. The function $f[ y ]$ satisfies $f[ 0 ] = c$ and $f[ y ] \sim \sqrt{b} / y$ 
when $y$ is sufficiently large. Figure 3 shows the relation between 
$\left \langle w^2_{~} \right \rangle$ and $\sqrt{\Delta} \, 
\lambda ( L + 1 )^2_{~}$ for $\Delta = 0.5$ and $\Delta = 1$. The plot supports 
the presence of the scaling function $f[ y ]$.
In the same manner, it is expected that the energy correction 
$\Delta_L^{~}( \lambda ) - \Delta$ satisfies the scaling relation
\begin{equation}
( L + 1 )^2_{~} \left( \Delta_L^{~}( \lambda ) - \Delta \right)
= g\left[ \sqrt{\Delta} \, \lambda ( L + 1 )^2_{~} \right] \, ,
\end{equation}
where $g[ 0 ] = \pi^2_{~}$ and $g[ y ] \sim \sqrt{b} y$ when $y$ is sufficiently 
large. Figure 4 shows the relation between $( L + 1 )^2_{~} 
\left( \Delta_L^{~}( \lambda ) - \Delta \right)$ and $\sqrt{\Delta} \, \lambda 
( L + 1 )^2_{~}$. Clearly, there is a scaling function $g[ y ]$ also for
$\Delta_L^{~}( \lambda ) - \Delta$. In summary, 
an advantage of hyperbolic deformation is that the effect of the system boundary disappears when 
$\sqrt{\Delta} \, \lambda ( L + 1 )^2_{~}$ is sufficiently large. This property 
would make the numerical determination of $\Delta$ easier.

\begin{figure}
\begin{center} \includegraphics[width=70mm]{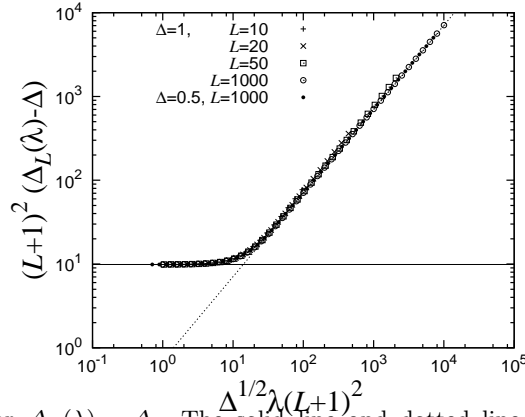} \end{center}
\caption{
Scaling plot for $\Delta_{L}^{~}(\lambda) - \Delta$. 
The solid line and dotted line, respectively, mean 
$(L+1)^2_{~}(\Delta_{L}^{~}(\lambda) - \Delta ) = \pi^2_{~}$ and 
$(L+1)^2_{~}(\Delta_{L}^{~}(\lambda) - \Delta ) = \sqrt{b} \, y$.
}
\end{figure}

\section{Conclusion and discussion}

We have introduced the hyperbolic deformation to 1D quantum lattice models. The 
quantum-classical correspondence suggests that the ground state is not strongly 
modified by the effect of deformation, while the excited quasi particle is weakly confined
around the center of the system. The mean-square width of the quasi-particle distribution
is calculated for a particle in the 1D lattice. As a result, we have obtained the 
scaling relation Eq.~(3$\cdot$16) for the finite-size correction to the excitation gap. 

\begin{figure}
\begin{center} \includegraphics[width=70mm]{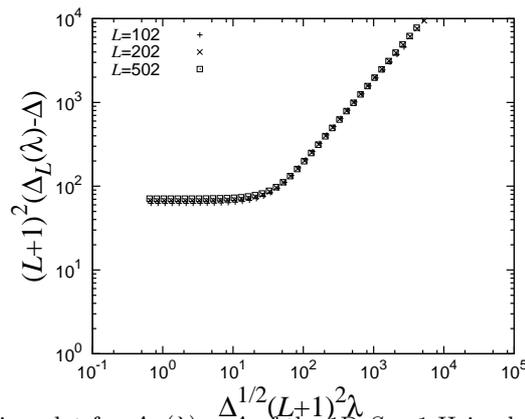} \end{center}
\caption{
Scaling plot for $\Delta_{L}^{~}(\lambda) - \Delta$ of the 1D $S = 1$ 
Heisenberg spin chain.
}
\end{figure}

As a confirmation of the scaling relation in realistic systems, we calculate the excitation energy 
of the hyperbolically deformed $S = 1$ antiferromagnetic Heisenberg spin chain,
where ${\hat h}_{j, j+1}^{~}$ in Eq.~(1$\cdot$3) is given 
by ${\bf S}_j^{~} \cdot {\bf S}_{j+1}^{~}$, using the DMRG method. 
Figure 5 shows the calculated result plotted in the same manner as in Fig.~4, where
the best fit is obtained when $\Delta$ is 0.410485. Numerical details and further
investigation would be presented elsewhere. It is thus expected that the scaling 
plot proposed in this article is useful for the numerical estimation of the excitation 
gap of various 1D correlated systems, such as the higher $S$ spin chains. 
In the case that the excited quasi particle occupies several 
sites on the lattice, one has to introduce an additional parameter $d$ and replace 
$( L + 1 )^2_{~}$ by $( L + d )^2_{~}$ to reduce higher-order corrections.

\section*{Acknowledgments}

The authors would like to thank Koichi Okunishi for valuable discussions. 
This work was partially supported by a Grant-in-Aid for JSPS Fellows, Grants-in-Aid 
from the Ministry of Education, Culture, Sports, Science and Technology (MEXT) 
(No.s~19540403, 20340096, and 22014012), and the 
Global COE Program (Core Research 
and Engineering of Advanced Materials-Interdisciplinary Education Center for 
Materials Science), MEXT, Japan.

\appendix
\section{Continuous limit of the one-particle Hamiltonian}

Let us check the continuum limit of Eq.~(\ref{sch_eq}).
Substituting the relations $x = a j$, $\lambda = a \nu$, and the correspondence
$\Psi_j^{~} = \Psi( a j ) = \Psi( x )$, we obtain
\begin{eqnarray}
E \,  \Psi( x ) & = & 
- t  \cosh \nu \bigl( x - \frac{a}{2} \bigr)
\cosh \nu \frac{a}{2}
\bigl[ \Psi( x + a ) + \Psi( x - a ) \bigr] \nonumber\\
&~&
- t  \sinh \nu \bigl( x - \frac{a}{2} \bigr)
\sinh \nu \frac{a}{2}
\bigl[ \Psi( x + a ) - \Psi( x - a ) \bigr]
- \mu \cosh \nu \bigl( x - \frac{a}{2} \bigr)
\Psi( x ) \nonumber\\
\label{sch_eq2}
\end{eqnarray}
after some algebra. 
Expressing the hopping amplitude as $t = \frac{\hbar^2_{~}}{2 m a^2_{~}}$, 
chemical potential as $\mu = - U - \frac{\hbar^2_{~}}{2m}\frac{\nu^2_{~}}{4} - 2 t$, 
and taking the limit $a \rightarrow 0$, we obtain a differential equation
\begin{equation}
E \, \Psi( x ) = \left[
- \frac{\hbar^2_{~}}{2m}  \frac{\partial}{\partial x}
\cosh \nu x \, \frac{\partial}{\partial x}
+ U \cosh \nu x \right] \Psi( x ) \, .
\label{sch_eq3}
\end{equation}
The first term in the r.h.s. is the deformed kinetic energy, and the second term 
is a kind of trapping potential when $U > 0$.
The Lagrangian that draws the above equation from the stationary condition is 
\begin{equation}
{\cal L}(\Psi^*,\partial_{\tt t}\Psi^*,
\partial_x\Psi^*,\Psi,\partial_{\tt t}\Psi,
\partial_x\Psi)
=
\Psi^*_{~} \frac{\partial}
{\partial {\tt t}} \Psi +
\cosh \nu x \left[
\frac{\hbar^2_{~}}{2m} \frac{\partial \Psi^*_{~}}
{\partial x} \frac{\partial \Psi}{\partial x} +
U \Psi^*_{~} \Psi
\right] \, ,
\label{lagrangian}
\end{equation}
where we have introduced the letter ${\tt t}$ for the imaginary-time variable, and where we have used the unit that satisfies $\hbar = 1$.
Note that the timelike variable $\tau$ in Eq.~(2.1) is related to ${\tt t}$ by the relation $( \cosh \nu x ) \, d {\tt t} = d \tau$, and in the $x$-$\tau$ plane, the Lagrangian can be represented as
\begin{equation}
{\cal L}'(\Psi^*,\partial_{\tau}\Psi^*,
\partial_x\Psi^*,\Psi,\partial_{\tau}\Psi,
\partial_x\Psi) = 
\cosh \nu x \left[
\Psi^*_{~} \frac{\partial}{\partial \tau} \Psi +
\frac{\hbar^2_{~}}{2m} \frac{\partial \Psi^*_{~}}
{\partial x} \frac{\partial \Psi}{\partial x} +
U \Psi^*_{~} \Psi
\right]
\, .
\label{lagrangian2}
\end{equation}
The action of the system is therefore written as
\begin{equation}
\int {\cal L}'(\Psi^*,\partial_{\tau}\Psi^*,
\partial_x\Psi^*,\Psi,\partial_{\tau}\Psi,
\partial_x\Psi) \, d\tau dx 
= \int
\left[\Psi^*_{~} \frac{\partial}{\partial
\tau} \Psi + \hat{h}(x) \right]
(\cosh \nu x) \, d\tau dx \, ,
\end{equation}
where the local Hamiltonian ${\hat h}( x )$ can be simply represented as
\begin{equation}
\hat{h}(x)=
\frac{\hbar^2_{~}}{2m} \frac{\partial \Psi^*_{~}}
{\partial x}
\frac{\partial \Psi}{\partial x} + U \Psi^*_{~}
\Psi
\end{equation}
under this coordinate.


\begin{thebibliography}{99}
\bibitem{White1} S.~R.~White, Phys. Rev. Lett. {\bf 69} (1992), 2863.
\bibitem{White2} S.~R.~White, Phys. Rev. B {\bf 48} (1993), 10345. \label{White2}
\bibitem{Peschel} I.~Peschel, X.~Wang, M.~Kaulke and K.~Hallberg (Eds.), 
{\it Density-matrix renormalization, a new numerical method in physics}, 
Lecture Notes in Physics (Springer, Berlin, 1999).
\bibitem{Schollwoeck} U.~Schollw\"{o}ck, Rev. Mod. Phys. {\bf 77} (2005), 259.
\bibitem{FSS1} M.~E.~Fisher, in {\it Proc. Int. School of Physics `Enrico Fermi',} ed. 
M.~S.~Green, (Academic Press, New York, 1971). Vol. {\bf 51}, 1.
\bibitem{FSS2} M.~N.~Barber, in {\it Phase transitions and critical phenomena,} ed. 
C.~Domb and J.~L.~Lebowitz, (Academic Press, New York, 1983), Vol. {\bf 8}, 146 
and references therein. 
\bibitem{vekic:PRL71} M.~Veki\'{c} and S.~R.~White, Phys. Rev. Lett. {\bf 71} (1993), 4283.
\bibitem{vekic:PRB53} M.~Veki\'{c} and S.~R.~White, Phys. Rev. B {\bf 53} (1996), 14552.
\bibitem{Andrej:PTP122} A.~Gendiar, R.~Krcmar and T.~Nishino, Prog. Theor. Phys. 
{\bf 122} (2009), 953;  Prog. Theor. Phys. {\bf 123} (2010), 393.
\bibitem{Huse} S.~R.~White and D.A.~Huse, Phys. Rev. B {\bf 48} (1993), 3844.
\bibitem{Schollwock:PRB54} U.~Schollw{\"o}ck, O.~Golinelli and T.~Jolic{\oe}ur, Phys. Rev. 
B {\bf 54} (1996), 4038.
\bibitem{Nakano} H.~Nakano and A.~Terai, J. Phys. Soc. Jpn. {\bf 78} (2009), 014003. 
\bibitem{Ueda} H.~Ueda and T.~Nishino, J. Phys. Soc. Jpn. {\bf 78} (2009), 014001.
\bibitem{Kanamori} J.~Kanamori, Prog. Theor. Phys. {\bf 3}0 (1963), 275.
\bibitem{Gutzwiller} M.~C.~Gutzwiller, Phys. Rev. Lett. {\bf 10} (1963), 275; Phys. Rev. {\bf 134} (1965), A1726.
\bibitem{Hubbard} J.~Hubbard, Proc. Roy. Soc. A {\bf 281} (1964), 401.
\bibitem{Trotter} H.~F.~Trotter, Proc. Am. Math. Soc. {\bf 10} (1959), 545.
\bibitem{Suzuki} M.~Suzuki, Prog. Theor. Phys. {\bf 56} (1976), 1454.  
\bibitem{Baxter} R.~J.~Baxter, {\it Exactly solved models in statistical mechanics} 
(Academic Press, London, 1982).
\bibitem{iharagi} R.~Krcmar, T.~Iharagi, A.~Gendiar and T.~Nishino, Phys. Rev. E {\bf 78} (2008), 061119.
\end{thebibliography}
\end{document}